\documentclass[prb,11pt, nofootinbib]{revtex4-1}
\usepackage{amsmath}    % need for subequations
\usepackage{graphicx}   % need for figures
\usepackage{verbatim}   % useful for program listings
\usepackage{hyperref}   % use for hypertext links, including those to external documents and URLs
\begin{document}

\title{Evidence of electric breakdown induced by bubbles in liquid argon\footnote{Based on a talk given at the High Voltage in Noble Liquids (HVNL13) Workshop, FNAL, 8-9 November 2013}}
\author{F. Bay}
\affiliation{ETH Zurich - Institute for Particle Physics, Otto-Stern-Weg 5, 8093 Zurich (Switzerland)}
\author{C. Cantini}
\affiliation{ETH Zurich - Institute for Particle Physics, Otto-Stern-Weg 5, 8093 Zurich (Switzerland)}
\author{S. Murphy}
\affiliation{ETH Zurich - Institute for Particle Physics, Otto-Stern-Weg 5, 8093 Zurich (Switzerland)}
\author{F. Resnati}\email{filippo.resnati@cern.ch}
\affiliation{ETH Zurich - Institute for Particle Physics, Otto-Stern-Weg 5, 8093 Zurich (Switzerland)}
\author{A. Rubbia}
\affiliation{ETH Zurich - Institute for Particle Physics, Otto-Stern-Weg 5, 8093 Zurich (Switzerland)}
\author{F. Sergiampietri}
\affiliation{ETH Zurich - Institute for Particle Physics, Otto-Stern-Weg 5, 8093 Zurich (Switzerland)}
\author{S. Wu}
\affiliation{ETH Zurich - Institute for Particle Physics, Otto-Stern-Weg 5, 8093 Zurich (Switzerland)}
\date{13 January 2014}

\begin{abstract}
We report on the results of a high voltage test in liquid argon in order to measure its dielectric rigidity.
Under stable conditions and below the boiling point, liquid argon was found to sustain a uniform electric field of 100~kV/cm, applied in a region of 20~cm$^2$ area across 1~cm thick gap.
When the liquid is boiling, breakdowns may occur at electric fields as low as 40~kV/cm.
This test is one of the R\&D efforts towards the Giant Liquid Argon Charge Imaging ExpeRiment (GLACIER) as proposed Liquid Argon Time Projection Chamber (LAr TPC) for the LBNO observatory for neutrino physics, astrophysics and nucleon decay searches.
\end{abstract}

\maketitle

\section{Introduction}
Now that the $\sin^2(2\theta_{13})$ has been measured~\cite{T2K:2011, DB:2012}, the future long-baseline neutrino experiments are expected to have significant discovery potential for CP violation, and the capability to establish the neutrino mass hierarchy.
The LBNO experiment~\cite{Stahl:2012} aims at very large statistics, excellent background rejection and good energy resolution in order to measure precisely the oscillation pattern as a function of the neutrino energy.
The Giant Liquid Argon Charge Imaging ExpeRiment (GLACIER) is a charge imaging detector which allows to reconstruct tracks in three dimensions and their locally deposited energy.
It provides high efficiency for $\nu_e$ charged current interactions with high rejection power against $\nu_\mu$ neutral and charged currents backgrounds in the GeV and multi-GeV region.
The fine spatial granularity guarantees a good e$^-$/$\pi^0$ separation.
Moreover, a good particle identification between charged K, $\mu$/$\pi$ and proton can be achieved from the local measure of the stopping power.
A key and innovative feature is the double phase (liquid-vapour) operation, which permits the amplification of the signal by means of charge avalanche in the vapour, yielding a larger signal to noise ratio and an overall better image quality~\cite{Cantini:2013, Badertscher:2013, Badertscher:2011a, Badertscher:2011b, Badertscher:2010, Badertscher:2008}.
Several technological challenges need to be addressed, the high voltage (HV) being a critical one.
The assumed electron drift length is 20~m, with an electric field of 1~kV/cm, corresponding to the path after which the spread due to the electron diffusion becomes larger than the This requires a voltage at the cathode of 2~MV (negative).
In the current design\cite{LAGUNA_LBNO}, a distance of 1.5~m is foreseen between the cathode structure and the bottom of the tank at ground, giving an average electric field of 13.3~kV/cm.
In the vicinity of the cathode tubes the electric field reaches 50~kV/cm over distances of the order of centimetre.
Therefore, we decided to perform a dedicated test to measure the maximum electric field (up to 100~kV over 1~cm) that the liquid argon can sustain.
The setup was operated for the first time in December 2013.
The description is given in section~\ref{sec:apparatus}, and we report on the results obtained in section~\ref{sec:test}.

\section{The apparatus}
\label{sec:apparatus}
\begin{figure}[htbp]
\centering
\includegraphics[width=0.9\textwidth]{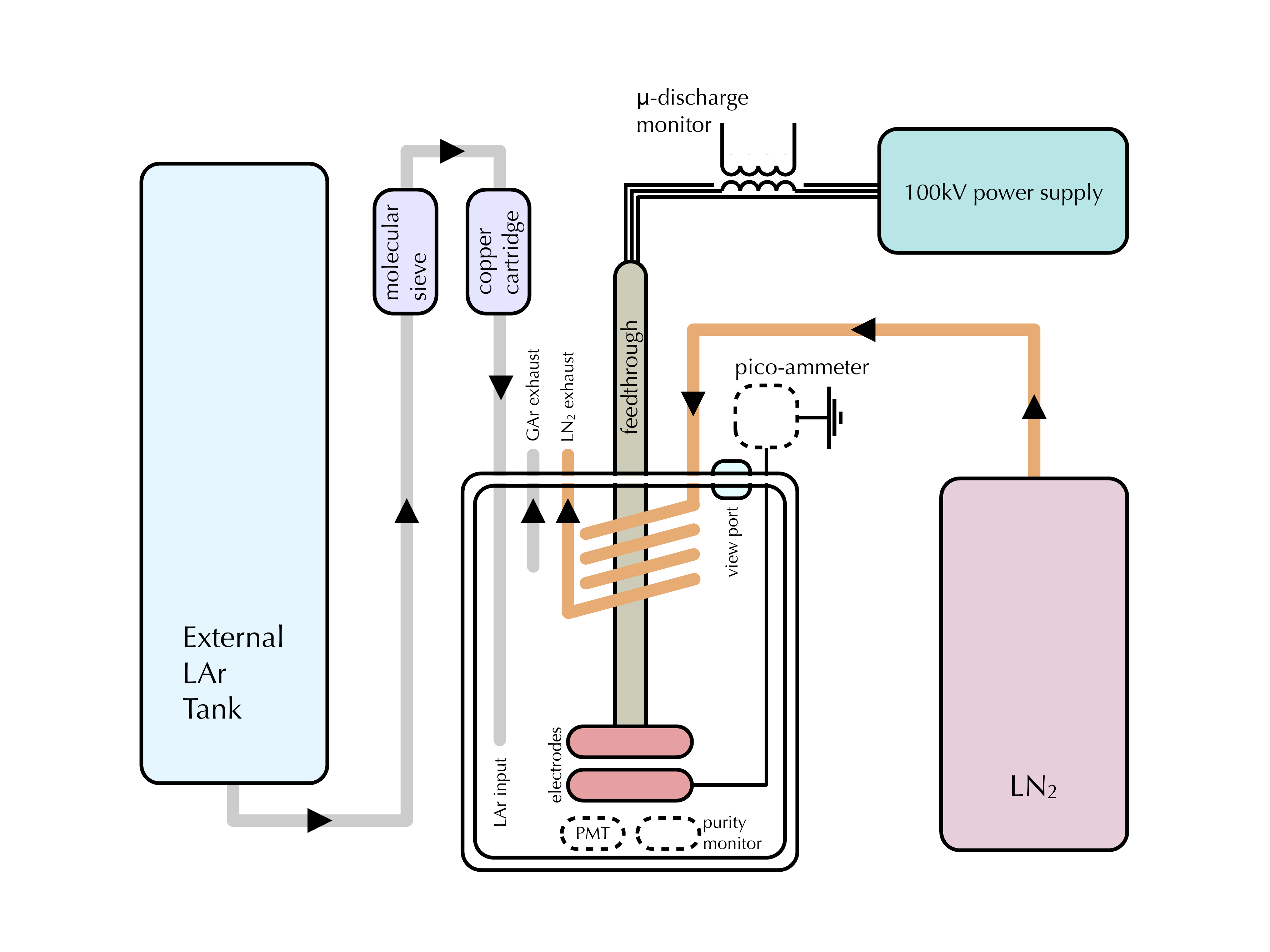}
\caption{Schematic representation of the apparatus. See text for the description.}
\label{fig:scheme}
\end{figure}

Figure~\ref{fig:scheme} shows the schematic representation of the test setup, which consists of a vacuum insulated dewar hosting a high voltage feedthrough and a couple of electrodes, in between which the high electric field is generated.

During operation, the dewar is filled with liquid argon purified through a molecular sieve (ZEOCHEM Z3-06), which blocks the water molecules, and a custom-made copper cartridge, which absorbs oxygen molecules.
Before the filling, the vessel is evacuated to residual pressure lower than $10^{-4}$~mbar in order to remove air traces, favour the outgassing of the materials in the dewar, and check the absence of leaks towards the atmosphere.
During the filling, the argon {\it boil off} is exhausted to control the pressure in the dewar.
The pressure is always kept at least 100~mbar above the atmospheric pressure, so that air contaminations are minimised.
Once the detector is completely full, the exhaust is closed, and the liquid argon is kept cold by means of liquid nitrogen flowing into a serpentine, which acts as heat exchanger that condenses the argon {\it boil off}.
The liquid argon level is visually checked through a vacuum sealed viewport.

\begin{figure}[htbp]
\centering
\includegraphics[width=0.9\textwidth]{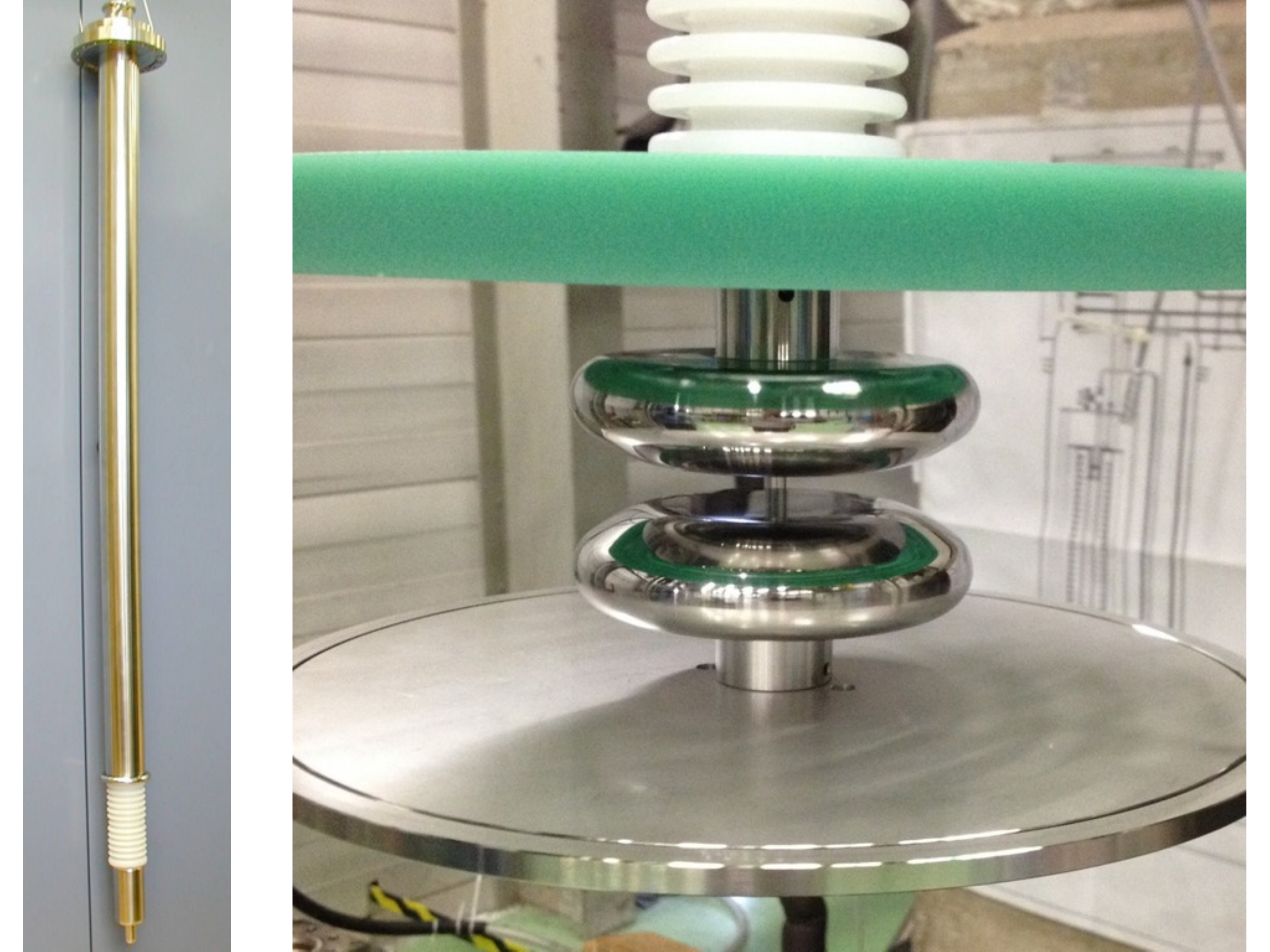}
\caption{Left: Image of the High voltage feedthrough. Right: figure of the electrodes structure.}
\label{fig:feedthrough}
\end{figure}

The voltage is provided by a 100~kV power supply\cite{heinzinger} though a HV coaxial cable modified to inductively couple the HV wire to an oscilloscope via a 1:1 transformer.
When a pulsed current flows through the cable, it is detected on the scope.
This sensor is used to monitor the frequency of small discharges (not necessarily happening between the electrodes, e.g.\ the feedthrough and the cable may have leakage currents).
It is also used to monitor the current delivered by the power supply during the charging up.
The cable enters into a custom-made HV feedthrough (see figure~\ref{fig:feedthrough} left) analogous to the one developed by ICARUS~\cite{ICARUS:2004}.
It is vacuum tight and designed to sustain voltages larger than 150~kV.

\begin{figure}[htbp]
\centering
\includegraphics[width=0.45\textwidth]{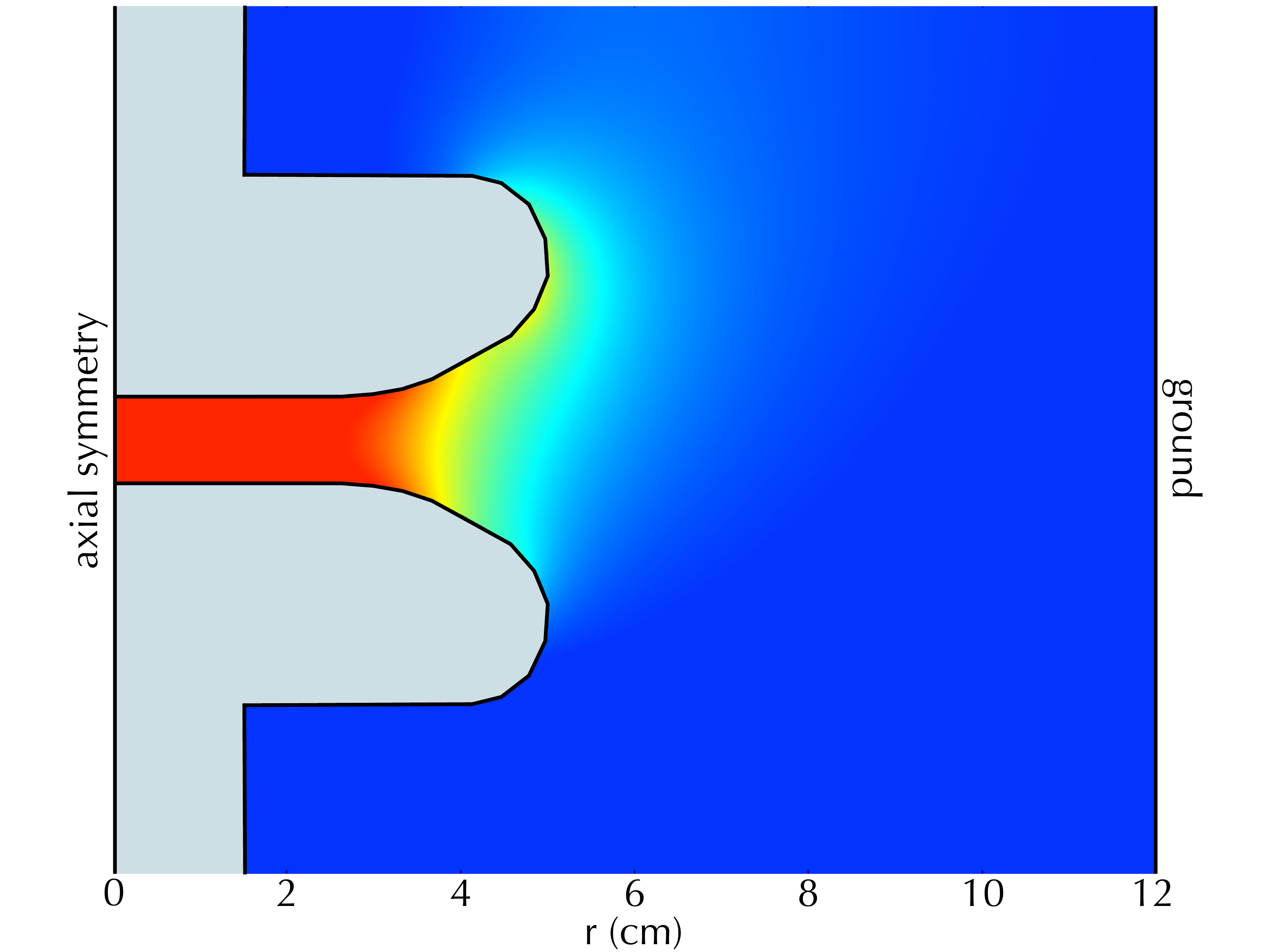}
\includegraphics[width=0.45\textwidth]{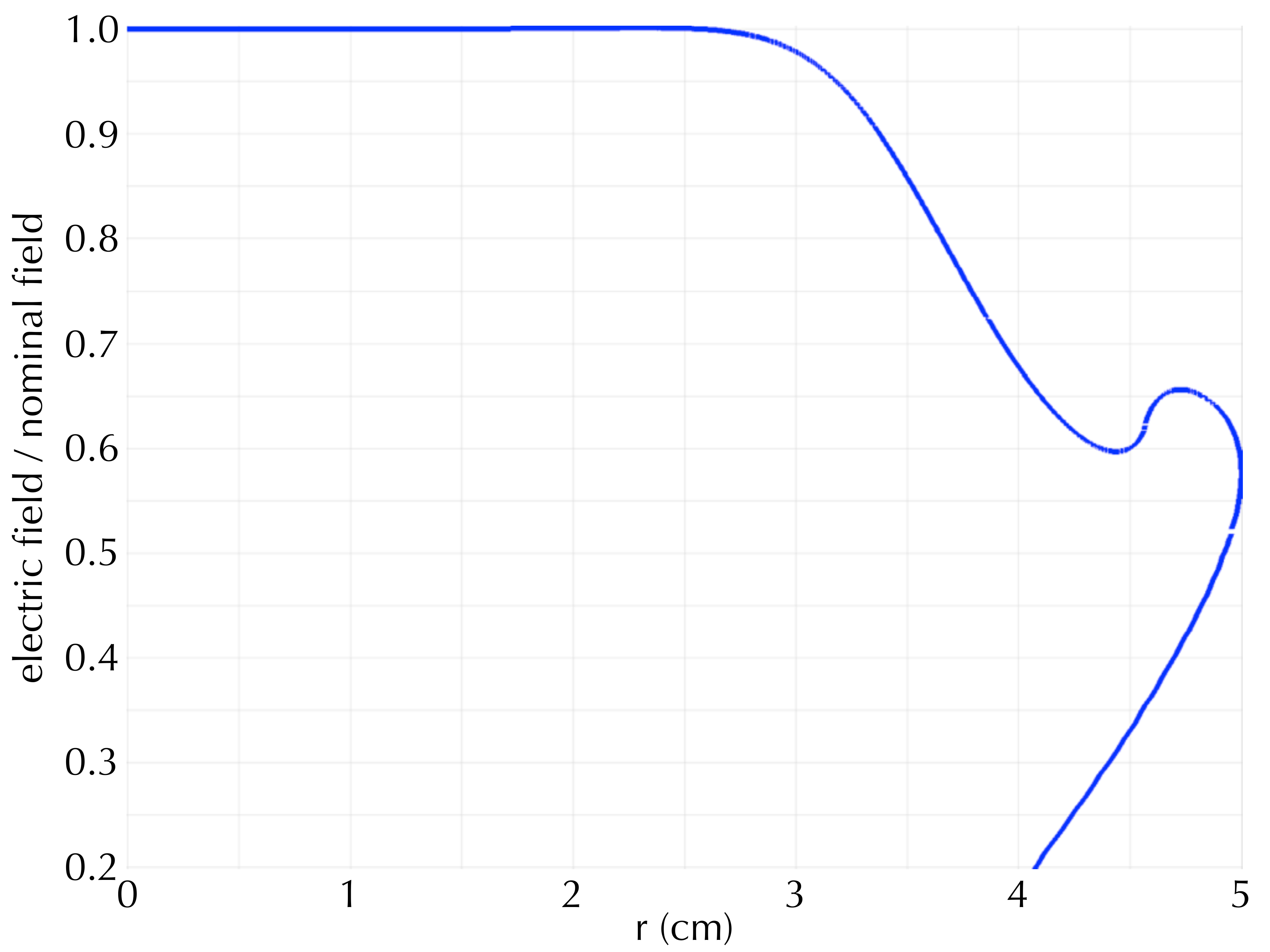}
\caption{Left: Computed electric field. The cross sections of the electrodes are shown in grey, and the colour pattern is proportional to the absolute value of the electric field. The electric field is essentially uniform in the central region. Right: Computed electric field on the profile of the top electrode as a function of the radius. The largest field is attained between the electrodes.}
\label{fig:field}
\end{figure}

High electric fields can be achieved with low potentials and electrodes with small curvature radii, but, since the breakdown is a random process, we believe it is important to test a sizeable region of the electrodes.
For these reasons, we designed a system that provides a uniform electric field over 20~cm$^2$ area.
A picture of the electrodes structure is shown in figure~\ref{fig:feedthrough} right.
The two 10~cm diameter electrodes have the same shape and are facing each other at a distance of 1~cm.
The top electrode is connected to the live contact of the HV feedthrough, and the bottom one is connected to ground through the vessel.
The electrodes, made out of mechanically polished stainless steel, are shaped according to the Rogowski profile~\cite{Rogowski:1923} that guarantees that the highest electric field is almost uniform (in a region of about 5~cm in diameter), and confined in between the two electrodes, as shown in figure~\ref{fig:field}.
The left image shows, in cylindrical coordinates, the absolute value of the electric field in the vicinity of the electrodes, computed with COMSOL\cite{comsol}.
On the right the electric field along the profile of the top electrode as a function of the radius is shown.

The two electrodes form a standalone structure, that is assembled first and then mounted.
By construction, the structure ensures the parallelism of the electrodes when cooled down to the liquid argon temperature.
The shrinkage of the materials in cold is computed to affect the distance between the electrodes less than 1\%.

\section{Results}
\label{sec:test}
In December 2013 the setup was operated for the first time.
The goal of the first test was to commission the setup, and to check if a field of 100~kV/cm can be reached in stable conditions.
The operation with the high voltage power supply switched on lasted about 4~hours.

With the liquid argon temperature below the boiling point at a given pressure, a voltage of -100~kV was applied to the top electrode.
This value was limited by the maximum voltage of the power supply.
This configuration corresponds to a uniform electric field of 100~kV/cm in a region of about 20~cm$^2$ area between the electrodes.
Several cycles of discharging and charging up of the power supply were performed.
The system could also be stressed several times by ramping up the voltage from 0~V to -100~kV in about 20~s without provoking any breakdown.

A completely different behaviour was observed with boiling argon.
We could cause several breakdowns between the electrodes at fields as low as 40~kV/cm.
The stillness of the liquid argon was controlled by varying the pressure of the argon vapour and was monitored visually by looking through the viewport.
The pressure was regulated by acting on the flow of the liquid nitrogen passing through the heat exchanger.
In fact, the thermal inertia of the liquid argon bulk makes temperature variations very slow, hence increasing the cooling power translates in an rapid decrease of the vapour pressure.
%The behaviour of the liquid argon could be monitored visually through the viewport.
When the pressure is above the boiling one, the liquid argon surface becomes flat and still.
On the contrary, the more the pressure decreases below the boiling point, the more the liquid argon boils.

Above about 1000~mbar the argon was not boiling, and 100~kV/cm could be achieved.
Around 930~mbar breakdowns occurred at 70~kV/cm, and below 880~mbar at 40kV/cm.
Since argon gas has a much lower dielectric rigidity compared to the liquid, we interpret this behaviour as the evidence that breakdown can be triggered by bubbles in the liquid.

\section{Conclusions and outlook}
The dielectric rigidity of liquid argon with temperature below the boiling point is larger than 100~kV/cm.
The occurrence of breakdowns was found to be significantly sensitive to the presence of bubbles: in boiling liquid argon the breakdown field is as low as 40~kV/cm.
We believe that this is due to the bubbles themselves that convey the breakdown forming a preferred channel for the discharge development, because of the much lower dielectric rigidity of the argon gas.
We intend to repeat the test to quantitatively measure the correlation between the bubble formation and the breakdowns.
%The electrode distance may be decreased to increase the maximum achievable electric field, given the actual HV power supply.
We plan also to test larger electrode distances with larger voltages and measure the dielectric rigidity of liquid argon as a function of the argon purity and electrode distance.
The setup will be upgraded installing a PhotoMultiplier Tube (PMT), a liquid argon purity monitor and a pico-ammeter, represented with dashed lines in figure~\ref{fig:scheme}.
The main purpose of the PMT is to establish the presence of luminescent precursors of the breakdown in liquid.
The purity monitor will consist of a small TPC to measure, with cosmic muons, the lifetime of the drifting ionisation electrons.
Since this setup will also be used to characterise dielectric materials in presence of high electric fields and in cryogenic conditions, a pico-ammeter will be used to measure the current flowing through the materials placed between the two electrodes, in order to measure their volumetric impedance and the surface currents.

\end{document}